\documentstyle[12pt,epic,eepic]{ioplppt}
\begin{document}
\jl{1}
\title{Low temperature series expansions for the square lattice 
Ising model with spin $S > 1$.}[Series expansions for the $S>1$
Ising model]

\author{I Jensen\dag\ftnote{3}{E-mail address: iwan@maths.mu.oz.au}, 
A J Guttmann\dag\ftnote{4}{E-mail address: tonyg@maths.mu.oz.au} and 
I G Enting\ddag\ftnote{5}{E-mail address: ige@dar.csiro.au}}
\address{\dag\ Department of Mathematics,
The University of Melbourne, Parkville, Victoria 3052, Australia.}

\address{\ddag\ CSIRO, Division of Atmospheric Research,
Mordialloc, Victoria 3195, Australia.}

\begin{abstract}
We derive low-temperature series (in the variable $u = \exp[-\beta J/S^2]$)
for the spontaneous magnetisation, susceptibility and specific heat
of the spin-$S$ Ising model on the square lattice for $S=\frac32$,
2, $\frac52$, and 3. We determine the 
location of the physical critical point and non-physical singularities.
The number of non-physical singularities closer to the origin than
the physical critical point grows quite rapidly with $S$. The critical 
exponents at the singularities which are closest to the origin and for
which we have reasonably accurate estimates are independent of $S$. 
Due to the many non-physical singularities, the estimates for the 
physical critical point and exponents are poor for higher values
of $S$, though consistent with universality. 
\end{abstract}

\pacs{05.50.+q, 05.70.Jk, 75.10.H}
\maketitle

\section{Introduction}

In an earlier paper \cite{enting94} we presented low-temperature 
series for the spontaneous magnetisation, susceptibility and 
specific heat of the spin-1 Ising model on the square lattice. In 
this paper we extend this work to higher spin values ($S=\frac32$, 
2, $\frac52$, and 3). 
From general theoretical considerations, in particular renormalization
group theory, it is expected that the critical exponents (at the
physical singularity) depend only upon the dimensionality of the
lattice and on the symmetry of the ordered state, and thus does not
vary with spin magnitude.
Numerical work on the Ising model with $S>1$ is quite sparse and
little has been published since the mid-70's. Low-temperature
expansions were obtained by Fox and Guttmann \cite{fox} for
$S=1$ and $S=\frac32$ for various two and three dimensional lattices.
High-temperature expansions have been reported by a number of authors
\cite{saul,camp,nickel} who mainly focused on three dimensional 
lattices. Generally the numerical work has confirmed spin independence.
Recently, Matveev and Shrock \cite{MSISzeros}
studied the distribution of zeros of the partition function of 
the square lattice Ising model for $S= 1,\; \frac32$, and 2. 
While the physical critical behaviour of the spin-$S$ Ising  model is 
fairly well understood, little is known about the non-physical 
singularities. One major reason for seeking more knowledge about
the complex-temperature behaviour is the hope that this will help
in the search for exact expressions for thermodynamic quantities
which have not yet been calculated exactly. 

\section{Low temperature series expansions}

The Hamiltonian defining the spin-S Ising model in a 
homogeneous magnetic field $h$ may be written:

\begin{equation}\label{eq:IsHam}
{\cal H} = \frac{J}{S^2}\sum_{\langle ij \rangle} (S^2-\sigma_i \sigma_j) +
           \frac{h}{S}\sum_{i}(S-\sigma_i)
\end{equation}
where the spin variables $\sigma_i$ may take the $(2S+1)$ values
$\sigma_i = S,S-1,\ldots,-S$. The first sum runs over all nearest 
neighbour pairs and the second sum over all sites. The constants
are chosen so the ground state ($\sigma_i = S\; \forall i$)
has zero energy. The low-temperature expansion, as described by 
Sykes and Gaunt \cite{sykes73}, is based on perturbations from the 
ground state. The expansion is expressed in terms of the 
low-temperature variable
$u = \exp(-\beta J/S^2)$ and the field variable 
$\mu = \exp(-\beta h/S)$, where $\beta = 1/kT$. The expansion 
of the partition function in powers of $u$ may be expressed as

\begin{equation}
Z = \sum_{k=0}^{\infty} u^{k}\Psi_{k}(\mu)
\end{equation}
where $\Psi_{k}(\mu)$ are polynomials in $\mu$. It is more
convenient to express the field dependence in terms of the
variable $x=1-\mu$ 

\begin{equation}
Z = \sum_{k=0}^{\infty} x^{k}Z_{k}(u).
\end{equation}

Using the standard definitions, we find the spontaneous magnetisation 

\begin{equation}
M(u) = M(0) + \frac{1}{\beta} \left. 
       \frac{\partial \ln Z}{\partial h}\right|_{h=0}
     = S+ Z_{1}(u)/Z_{0}(u),
\end{equation}
since $x=0$ in zero field. For the zero-field susceptibility we find

\begin{equation}\fl
\chi(u) = \left. \frac{\partial M}{\partial h}\right|_{h=0} =
\frac{\partial}{\partial h}\left. \left(
Z^{-1}\frac{\partial Z}{\partial h}\right)\right|_{h=0} =
\beta/S^2 \left[2\frac{Z_{2}(u)}{Z_{0}(u)}-\frac{Z_{1}(u)}{Z_{0}(u)}-
\left( \frac{Z_{1}(u)}{Z_{0}(u)} \right)^{2}\right].
\end{equation}

The specific heat series is derived from the zero field partition
function (via the internal energy 
$U= -(\partial/\partial \beta) \ln Z_{0}$),

\begin{equation}
C_{v}(u) = \frac{\partial U}{\partial T} = 
\beta^2\frac{\partial^{2}}{\partial \beta^{2}}\ln Z_{0}=
(\beta J/S^2)^2\left( u\frac{\mbox{d}}{\mbox{d}u} \right)^{2} 
\ln Z_{0}(u).
\end{equation}

Thus in order to calculate the specific heat, spontaneous magnetisation
and susceptibility one need only calculate the first three moments
(with respect to $x$), $Z_k(u)$ for $k \leq 2$, of the partition function.
These moments are most efficiently evaluated
using the finite lattice method. The algorithm was described in an
earlier paper \cite{enting94}. For our present purpose it suffices
to note that the infinite lattice partition function $Z$ can be 
approximated by a product of partition functions $Z_{mn}$ on 
{\em finite} ($m \times n$) lattices,

\begin{equation}
Z \approx \prod_{m,n} Z_{mn}^{a_{mn}} 
\mbox{\hspace{2cm} with } m \leq n \mbox{ and } m+n \leq r.
\end{equation}

The weights $a_{mn}$ were derived by Enting \cite{enting78}, and
are modified in the present algorithm to utilize the rotational 
symmetry of the square lattice. The number of terms derived correctly 
with the finite lattice method is given by the power of the 
lowest-order connected graph not contained in any of the rectangles 
considered. We use the {\em time-limited} version of the algorithm
\cite{enting94} in which the largest rectangles are determined by
a cut-off parameter $b_{\rm max}$, $m+n \leq r = 3b_{\rm max} +2$.
The simplest connected graphs not contained in such rectangles
are chains of $r$ sites all in the `$S-1$' state. From
\eref{eq:IsHam} we see that such chains give rise to 
terms of order $2(r+1)[S^2-S(S-1)]+(r-1)[S^2-(S-1)^2] = r(4S-1)+1$,
from the $2(r+1)$ interactions between
spins in states `$S$' and `$S-1$' and 
the $r-1$ interactions between spins both in state `$S-1$'.
For a given value of $b_{\rm max}$ the series expansion is thus correct 
to order $u^{(3b_{\rm max}+2)(4S-1)}$. In an earlier paper \cite{enting94}
we reported on the $S=1$ case where we went to $b_{\rm max}=8$ giving a 
series correct to $u^{78}$. We have since extended these series to $u^{113}$
using a more efficient parallel algorithm and a new extrapolation
procedure \cite{JGS1}. For the present
work we have calculated the series expansions for $S=\frac32$, 2, 
$\frac52$ and 3, deriving series correct to $u^{100}$ 
($b_{\rm max} = 6$) for $S=\frac32$, $u^{119}$ ($b_{\rm max} = 5$) 
for $S=2$, $u^{126}$ ($b_{\rm max} = 4$) for $S=\frac52$, and 
$u^{154}$ ($b_{\rm max} = 4$) for $S = 3$. 

\section{Analysis of the series}

The series for the spontaneous magnetisation, the susceptibility and
the specific heat of the spin-$S$ Ising model are expected to exhibit
critical behaviour of the forms

\begin{equation}
M(u) \sim \prod_j A_j(u_j-u)^{\beta_j}[1 + a_{j,1}(u_j-u) + 
                  a_{j,\Delta} (u_j-u)^{\Delta_j} + \ldots ],
\end{equation}

\begin{equation}
\chi(u) \sim \prod_j B_j(u_j-u)^{-\gamma_j '}[1+b_{j,1}(u_j-u) +
                     b_{j,\Delta} (u_j-u)^{\Delta_j} + \ldots ],
\end{equation}

\begin{equation}
C_v (u) \sim \prod_j C_j(u_j-u)^{-\alpha_j '}[1+c_{j,1}(u_j-u) + 
                     c_{j,\Delta} (u_j-u)^{\Delta_j} + \ldots ],
\end{equation}
where the terms involving $\Delta_j$ represent the leading 
non-analytic confluent singularity and the dots $\ldots$ represent 
higher order analytic and non-analytic confluent terms. By 
universality it is expected that the leading critical exponents at 
the physical singularity, $u_c$, equal those of the spin-$\frac12$ 
Ising model, i.e., $\beta=\frac18$, $\gamma ' = \frac74$ and 
$\alpha ' = 0$ (logarithmic divergence). 

We analysed the series using differential approximants (see
Ref.~\cite{serana} for a comprehensive review), which allows one
to locate the singularities and estimate the associated critical
exponents fairly accurately, even in cases such as these where there
are many singularities. We find that ordinary Dlog Pad\'{e} approximants
(first order homogeneous differential approximants) yield the most
accurate estimates for the physical singularity of
the magnetisation series, whereas first- and second-order inhomogeneous 
approximants are required in order to analyse the susceptibility and 
specific heat series. Here it suffices to say that a $K$th-order
differential approximant to a function $f$ is formed by matching the
first series coefficients to an inhomogeneous differential equation
of the form (see \cite{serana} for details)

\begin{equation}
\sum_{i=0}^K Q_{i}(x)(x\frac{\mbox{d}}{\mbox{d}x})^i f(x) = P(x) 
\end{equation}
where $Q_i$ and $P$ are polynomials of order $N_i$ and $L$,
respectively. First and second order approximants are denoted by
$[L/N_0;N_1]$ and $[L/N_0;N_1;N_2]$, respectively.

\subsection{The physical singularity}

In this section we focus on the behaviour at the physical critical
point. First we give a somewhat detailed summary of the analysis of 
the spin-$\frac32$ series so as to introduce the various techniques 
and approximation procedures that we have applied in the analysis. 
Generally the estimates for the critical parameters at the physical
singularity are quite poor because the series have many non-physical
singularities closer to the origin and even for the spin-1 series 
\cite{enting94,JGS1} the convergence of the estimates to the true values
of the critical parameters is very slow. We see no evidence
that the critical exponents of spin-$S$ Ising model aren't
in agreement with the universality hypothesis. Under this assumption,
we have derived improved estimates for the location of the physical
critical point and the critical amplitudes.

In \tref{tab:sm32} we have listed the estimates for the physical
singularity and critical exponent for the spontaneous magnetisation
of the spin-$\frac32$ Ising model. The estimates were obtained from
homogeneous differential approximants (which are equivalent to
Dlog Pad\'e approximants). There is a quite substantial spread
among the various approximants with most approximants 
yielding estimates around $u_c \simeq 0.7380$ and $\beta \simeq 0.130$.
The estimates of $\beta$, while generally on the large side, are
consistent with expectations of universality which would indicate
that $\beta = \frac18$. If we assume this value to be exact we
see that the approximants (assuming a linear dependence of $\beta$
on $u_c$) would lead to $u_c \simeq 0.73775$.

\Table{\label{tab:sm32} Estimates for $u_c$ and $\beta$ for
the spin-$\frac32$ Ising model as obtained from $[N,M]$ homogeneous
first-order differential approximants.} 
\br
\multicolumn{1}{r}{N} &
\multicolumn{2}{c}{[N-1,N]} &
\multicolumn{2}{c}{[N,N]} &
\multicolumn{2}{c}{[N+1,N]}\\ \hline
\multicolumn{1}{r}{} &
\multicolumn{1}{c}{$u_c$} &
\multicolumn{1}{c}{$\beta$} &
\multicolumn{1}{c}{$u_c$} &
\multicolumn{1}{c}{$\beta$} &
\multicolumn{1}{c}{$u_c$} &
\multicolumn{1}{c}{$\beta$} \\ 
\mr
40 & 0.738148  & 0.1306  & 0.738167  & 0.1308  & 0.738049  & 0.1295  \\ 
41 & 0.738124  & 0.1303  & 0.738020  & 0.1291  & 0.738081  & 0.1298  \\ 
42 & 0.737908  & 0.1275  & 0.737948  & 0.1281  & 0.737125  & 0.1085  \\ 
43 & 0.737918  & 0.1277  & 0.738046  & 0.1294  & 0.738099  & 0.1300  \\ 
44 & 0.738128  & 0.1303  & 0.738105  & 0.1301  & 0.738098  & 0.1300  \\ 
45 & 0.738123  & 0.1303  & 0.738059  & 0.1296  & 0.740267  & 0.1038  \\ 
46 & 0.737958  & 0.1283  & 0.738135  & 0.1304  & 0.738140  & 0.1304  \\ 
47 & 0.738140  & 0.1304  & 0.738135  & 0.1304  & 0.738331  & 0.1317  \\ 
48 & 0.736928  & 0.1047  & 0.737705  & 0.1242  & 0.737673  & 0.1236  \\ 
49 & 0.737676  & 0.1236  & 0.737700  & 0.1241  & 0.737867  & 0.1271  \\ 
50 & 0.738187  & 0.1313  & 0.737810  & 0.1261  & & \\ 
\br
\endTable

In \tref{tab:chi32} and \tref{tab:sh32} we have listed estimates for the 
position of the physical singularities and critical exponents
of the series for susceptibility and specific heat of the spin-$\frac32$
model. Since the first non-zero term in these series is $u^6$ the 
estimates were obtained by analysing the series $\chi(u)/u^6$ and
$C_v (u)/u^6$. The estimates were obtained by averaging first order 
$[L/N;M]$ and second order $[L/N;M;M]$ inhomogeneous differential 
approximants with $|N-M| \leq 1$. For each order $L$ of the 
inhomogeneous polynomial we averaged over most approximants to the 
series which as a minimum used all the series terms up to the 
last 15 or so. Some approximants were excluded from the averages 
because the estimates were obviously spurious. Examples include the 
[47,48] and [46,45] approximants in \tref{tab:sm32}. The error
quoted for these estimates reflects the spread (basically one standard
deviation) among the approximants. Note that these error bounds should
{\em not} be viewed as a measure of the true error as they cannot include
possible systematic sources of error. While the estimates aren't very
good, we see that the estimates for $u_c$ are consistent with the 
value $u_c \simeq 0.73775$ obtained from the magnetisation series by
demanding $\beta = \frac18$ and that the exponent estimates are consistent
with universality expectations of $\gamma ' = 7/4$ and $\alpha ' =0$.

\Table{\label{tab:chi32} Estimates for $u_c$ and $\gamma '$ for
the spin-$\frac32$ Ising model as obtained from inhomogeneous
first- and second-order differential approximants.} 
\br
 \multicolumn{1}{r}{} &
 \multicolumn{2}{c}{1. order DA } &
 \multicolumn{2}{c}{2. order DA }\\ \hline
 \multicolumn{1}{r}{L} &
 \multicolumn{1}{c}{$u_c$} &
 \multicolumn{1}{c}{$\gamma'$} &
 \multicolumn{1}{c}{$u_c$} &
 \multicolumn{1}{c}{$\gamma'$} \\ \hline
 0 & 0.73787(40) & 1.848(63) & 0.73802(37) & 1.864(58)  \\
 1 & 0.73808(31) & 1.882(52) & 0.73810(26) & 1.868(49)  \\
 2 & 0.73800(19) & 1.864(34) & 0.73818(20) & 1.882(39)  \\
 3 & 0.73804(23) & 1.874(43) & 0.73804(33) & 1.848(72)  \\
 4 & 0.73792(48) & 1.82(10) & 0.73805(38) & 1.863(69)  \\ 
 5 & 0.73803(46) & 1.895(65) & 0.73808(25) & 1.852(69)  \\
 6 & 0.73787(53) & 1.839(99) & 0.73803(53) & 1.82(18)  \\ 
 7 & 0.73823(18) & 1.50(98) & 0.73792(51) & 1.80(12)  \\  
 8 & 0.73774(64) & 1.76(20) & 0.73808(31) & 1.861(62)  \\ 
\br
\endTable

\Table{\label{tab:sh32} Estimates for $u_c$ and $\alpha '$ for
the spin-$\frac32$ Ising model as obtained from inhomogeneous
first- and second-order differential approximants.} 
\br
 \multicolumn{1}{r}{} &
 \multicolumn{2}{c}{1. order DA } &
 \multicolumn{2}{c}{2. order DA }\\ \hline
 \multicolumn{1}{r}{L} &
 \multicolumn{1}{c}{$u_c$} &
 \multicolumn{1}{c}{$\alpha'$} &
 \multicolumn{1}{c}{$u_c$} &
 \multicolumn{1}{c}{$\alpha'$} \\ \hline
 0 & 0.74062(88) & 0.343(16) & 0.7393(18) & 0.17(25)  \\  
 1 & 0.74030(88) & 0.320(80) & 0.7382(15) & 0.20(80)  \\  
 2 & 0.7397(20) & 0.24(32) & 0.7389(18) & 0.12(20)  \\ 
 3 & 0.7401(10) & 0.32(10) & 0.7384(16) & 0.07(23)  \\ 
 4 & 0.7370(28) & 0.06(71) & 0.7381(17) & 0.03(31)  \\ 
 5 & 0.7381(21) & 0.04(38) & 0.7378(21) & 0.05(48)  \\ 
 6 & 0.7373(25) & 0.24(61) & 0.7388(29) & 0.07(53)  \\ 
 7 & 0.7357(24) & 0.21(64) & 0.7381(28) & 0.33(90)  \\ 
 8 & 0.7356(24) & 0.25(68) & 0.7386(25) & 0.02(66)  \\ 
\br
\endTable

As for the critical exponents, it is obvious that the behaviour at
$u_c$ (except for $S=\frac12$ and 1) isn't represented very well by
the series. This discrepancy, which becomes more pronounced as 
$S$ increases, is hardly surprising given that the number of 
non-physical singularities within the physical disc increases
rapidly with spin magnitude (see following section for details). The
quite complicated singularity structure of the series simply tends
to obscure the behaviour at the physical singularity. This problem
is possibly further aggravated by the presence of confluent terms.
The only series which yields reasonably accurate estimates is the
magnetisation from which we estimate $\beta = 0.139(4)$, 0.138(5)
and 0.132(2) for $S=2,\; \frac52$ and 3, respectively. Again, the
quoted errors are merely a measure of the spread among the 
approximants rather than the true error. The differential 
approximant analysis of the higher $S$ series for the susceptibility
and specific heat yields little of value. Estimates for the 
critical exponent $\gamma '$ fluctuate wildly and lie somewhere
between 0.5 and 2 while generally favouring values below $7/4$. Similarly,
estimates for $\alpha '$ lie between -0.5 and 1. So while no
sensible estimates can be obtained there is no evidence
to suggest that the exponents are not consistent with universality.

While this situation is somewhat disappointing it is hardly surprising
in light of the behaviour of the spin-1 series, where our earlier
analysis showed a very slow convergence of estimates towards the
true values of the critical parameters \cite{enting94,JGS1}.
Though the order to which the higher spin-$S$ series are correct
exceeds that of the spin-1 series this is really just a consequence
of the definition of the expansion variable $u$. We would expect
the accuracy of estimates to depend not so much on the actual order of 
the series as much as on the maximal cut-off given by $b_{\rm max}$.
In essence the accuracy is determined by the number of distinct
graphs, consisting of spins flipped from the ground state (irrespective
of the actual value of the spins), that one has summed over. One
should therefore not expect more accurate estimates from the
higher spin-$S$ series than those one could have obtained by
truncating the spin-1 series at an order determined by the 
associated value of $b_{\rm max}$.

One may hope to obtain improved estimates for $u_c$ by raising the
relevant series to the power $1/\lambda$, where $\lambda$ is the 
expected leading critical exponent, and look for simple zeros and poles
of the resulting series. This procedure of biasing works quite well 
for the magnetisation and susceptibility series (it obviously cannot
be used for the specific heat series). 
It is well known that the analysis of series exhibiting a logarithmic 
divergence, as we expect of the specific heat series, is particularly 
difficult. A fairly simple way of circumventing these problems is
to study the derivative of the specific heat, d/d$u C_v (u)$. The 
series for this quantity should have a simple pole at $u_c$, a
situation much more amenable to analysis by either differential
approximants or even just ordinary Pad\'{e} approximants. This
approach does indeed confirm the logarithmic divergence at $u_c$,
though the evidence becomes rather circumstantial for higher
values of $S$. The estimates for $u_c$ derived in this fashion
are tabulated in  \tref{tab:biasps} and
were obtained by averaging ordinary $[N+K,N]$ Pad\'e approximants 
($K=0,\pm 1$) with $2N+K+15$ not less than the order
of the series. The error quoted for these estimates again 
merely reflects the spread among the approximants.

\Table{\label{tab:biasps} Biased estimates for the physical singularity.} 
\br
S & Magnetisation & Susceptibility & Specific Heat \\ 
\mr
$\frac32$ & 0.73774(2) & 0.7372(2) & 0.7379(5) \\
2   & 0.8293(2) & 0.8288(2) & 0.833(3) \\
$\frac52$ & 0.8795(3) & 0.881(3) & 0.882(2) \\
3  & 0.9107(4) & 0.914(1) & 0.905(4) \\
\br
\endTable

It is often
possible to find a transformation of variable which will map the
non-physical singularities outside the transformed physical disc.
One such transformation is given by $u=x/(2-x)$. Although the
series in the transformed variable have radii of convergence
determined by the physical singularity, this transformation
turns out to be of little use and does not allow us obtain better 
estimates for the critical parameters. This is probably because 
there still are singularities close to the physical disc and because
such singularity-moving transformations may introduce long-period
oscillations \cite{serana}. 

We have calculated the critical amplitudes using two different
methods, both of which are very simple and easy to implement.
In the first method, we note that if $f(u) \sim A(1-u/u_{c})^{-\lambda}$,
then it follows that
$(u_{c} -u)f^{1/\lambda}|_{u=u_{c}} \sim A^{1/\lambda}u_{c}$. So we
simply form the series for $g(u) = (u_{c} -u)f^{1/\lambda}$ and
evaluate Pad\'{e} approximants to this series at $u_{c}$. The result
is just $A^{1/\lambda}u_{c}$. This procedure works well for the
magnetisation and susceptibility series (it obviously cannot
be used to analyse the specific heat series). For the specific
heat series two different approaches have been used. In the
first approach we use the `trick' applied previously and look
at the derivative of the specific heat series for which the
above method should work with $\lambda =1$. In \tref{tab:ampl} we 
have listed the estimates for the critical amplitudes obtained in 
this fashion. As usual estimates for any given value of $u_c$
were obtained by averaging over many higher order approximants,
and the error-estimates in \tref{tab:ampl} reflect both the spread among
the various approximants as well as the depence on $u_c$.
In the second approach 
we start from $f(u) \sim A\ln (1-u/u_{c})$ and form the series
$g(u) = \exp (-f(u))$ which has a singularity at $u_c$ with 
exponent $A$. One virtue of this approach is that no prior estimate
of $u_c$ is needed. However, the spread among estimates from different
approximants is  very substantial though consistent with \tref{tab:ampl}.
Biasing the estimates at $u_c$ also confirms the value of the amplitude
though generally the spread is larger than for the first approach.
For the spin-3 susceptibility and specific heat series we could not obtain
reliable amplitude estimates since the spread tended to be larger 
than the average value and the poor estimate of $u_c$ leads to
even greater errors.

\Table{\label{tab:ampl} Estimates for the amplitudes
at the physical singularity.} 
\br
S & Magnetisation & Susceptibility & Specific Heat \\ 
\mr
$\frac32$ & 1.875(5) & 0.019(3) & 52(2) \\
2   & 2.57(2) & 0.0088(5) & 110(5) \\
$\frac52$ & 3.33(3) & 0.006(2) & 190(10) \\
3  & 4.10(5) & --- & --- \\
\br
\endTable

In the second method, proposed by Liu and Fisher \cite{LF}, one
starts from $f(u) \sim A(u)(1-u/u_{c})^{-\lambda}+B(u)$ and then
forms the auxiliary function  $g(u) = (1-u/u_{c})^{\lambda}f(u)
\sim A(u) + B(u)(1-u/u_{c})^{\lambda}$. Thus the required amplitude
is now the {\em background} term in $g(u)$, which can be obtained
from inhomogeneous differential approximants \cite{serana}.
This method can also be used to study the specific heat series. One
now starts from $f(u) \sim A(u)\ln (1-u/u_{c})+B(u)$ and then looks
at the auxiliary function $g(u) = f(u)/\ln (1-u/u_{c})$. As before
the amplitude can be obtained as the background term in $g(u)$.
This analysis yields amplitude estimates consistent with those
in \tref{tab:ampl}, but with larger error-bars.

In \tref{tab:ps} we have listed our final estimates for the physical 
singularities and the associated exponents and amplitudes. For the
estimates of the position of the physical singularities we have placed
most weight on the biased analysis of the magnetisation series.
In the spin-$\frac12$ case, $u_c$ and 
the exponents $\alpha '$ and $\beta$ and the amplitudes
$A_C$ and $A_M$ are known exactly due to the 
calculation of the free energy by Onsager \cite{onsager} and the
magnetisation by Yang \cite{yang}. The susceptibility amplitude
$A_{\chi}$ is known to very high precision \cite{IsSus}.
The spin-1 estimates are from \cite{JGS1}.

\fulltable{\label{tab:ps} The physical singularities 
and associated exponents and amplitudes.}
\br
S & $u_c$ & $\beta$ & $A_M$ & $\gamma '$ & $A_{\chi}$ & $\alpha '$ & $A_C$ \\ 
\mr
$\frac12$ & $3-2\sqrt{2}$ & $\frac18$ & 1.138789 & $\frac74$ 
  & 0.584850 & 0 & 5.40658 \\
1 & 0.5540653(5) & 0.12507(3) & 1.2083(2) & 1.750(1) & 0.0617(1) 
  &  0.0005(10) & 22.3(5)\\
$\frac32$ & 0.73775(15) & 0.128(3) & 1.875(15) & 1.85(15) &0.019(5) 
  & 0.0(3) & 52(4)  \\
2 & 0.8293(3) & 0.139(4) & 2.57(4) & ---  & 0.009(1) & --- & 110(10) \\
$\frac52$ & 0.8795(5) & 0.138(5) & 3.33(6) &  --- & 0.006(2) & --- & 190(20) \\
3   & 0.911(1) & 0.132(2) & 4.1(1) & --- & --- & --- & --- \\
\br
\endfulltable

\subsection{Non-physical singularities}

Except for $S=\frac12$ the series have a radius of covergence
smaller than $u_c$ due to singularities in the complex $u$-plane closer 
to the origin than the physical critical point. Since all the 
coefficients in the expansion are real, complex singularities always
come in pairs. The number of non-physical singularities appears
to increase quite dramatically with $S$ thus making it exceedingly
hard to locate them accurately for large $S$. 

In order to locate the non-physical singularities in a systematic
fashion we used the following procedure: We calculate all 
$[L/N;M]$ inhomogeneous first order differential approximants 
with $|N-M| \leq 1$ using all or almost all series terms 
for $10 \leq L \leq 16$. (We discard no more than the last 15-20
terms.) Each approximant yields $M$ possible singularities and 
associated exponents from the $M$ zeroes of $Q_1$ (many of these are 
of course not actual singularities of the series but merely 
spurious zeros of $Q_1$). Next we sort these `singularities' into
equivalence classes by the criterion that they lie at most a distance
$2^{-k}$ apart. An equivalence class is accepted as a singularity
if it contains more than $N_a$ approximants ($N_a$ can be adjusted but
we typically use a value around 2/3 of the total number
of approximants), and an estimate for the singularity and exponent
is obtained by averaging over the approximants (the spread among the
approximants is also calculated). This calculation is then repeated 
for $k-1$, $k-2$, $\ldots$ until a minimal value of 5 or so. 
To avoid outputting well converged singularities at every level, 
once an equivalence class has been accepted, the approximants 
which are members of it are removed, and the subsequent
analysis is carried out on the remaining data only. This procedure
is applied to each series in turn producing tables of possible
singularities. Next we look at these tables in order to identify
the true singularities. 

In \tref{tab:nps} we have listed the non-physical 
singularities that we believe to have been 
identified with some degree of certainty and accuracy. For higher
spin values several of these are marred by large error bounds and 
it is quite possible that we haven't been able to locate all 
non-physical singularities of the series, particularly for 
$S = \frac52$ and 3. First we accepted any singularity which
appeared in all the series at a reasonably early level, say $k \geq 10$.
These singularities are marked 1 in \tref{tab:nps} and all of them are
undoubtedly true singularities.
Singularities which appear for $k < 10$ are a lot more tricky
to deal with. Generally we also expect that a singularity which
appears for $k=8$ or 9 (or higher) in all series and for the majority
of values of $L$ is a true singularity of the series (these are marked 2
in \tref{tab:nps}). However,
we often find that some singularities appear for $k=8$ or higher in
some series but at lower values of $k$ all the way down to 5 in
other series, and it is not easy to determine which ones are true 
singularities and which ones are not. Those marked 3 appear in all
series and for all values of $L$ while those marked 4 appear in some
series for all $L$ but not neccesarily for all $L$ in other series.

\Table{\label{tab:nps} Non-physical singularities $u_s$
and associated exponents of the spin-$S$ series.}
\br
 & $u_s$ & $|u_s|/u_c$ & $\beta$ & $\gamma '$ & $\alpha '$ \\ 
\mr \multicolumn{6}{c}{Spin-1} \\ \mr
1 & -0.598550(5) & 1.08 & 0.1248(3) & 1.750(5) & 0.005(10) \\
1 & -0.3019395(5)$\pm 0.3787735(5)i$ & 0.87 & -0.1690(2) 
  & 1.1692(2) & 1.1693(3) \\
\mr \multicolumn{6}{c}{Spin-$\frac32$} \\ \mr
3 & 0.63(1)$\pm 0.45(1)i$ &1.05 & -1.8(5) & 2.7(5) & 2.4(6) \\
1 & 0.09477(2)$\pm 0.64117(5)i$ &0.88 & -0.174(5)& 1.185(5) & 1.185(1) \\
2 & -0.0654(5)$\pm 0.7113(4)i$ &0.97 &-0.18(3) & 1.21(2) & 1.22(3) \\
1 & -0.52924(2)$\pm 0.33797(2)i$ &0.85 &-0.177(5) & 1.184(5) & 1.188(5) \\
\mr \multicolumn{6}{c}{Spin-2} \\ \mr
2 & -0.842(5) &1.02 & 0.130(4) & 1.2(5) & 0.3(4) \\
1 & 0.3767(2)$\pm 0.6401(1)i$ & 0.90 & -0.16(3) & 1.19(1) & 1.19(3)\\
2 & 0.302(6)$\pm 0.727(8)$ &0.95 & --- & 1.3(4) & 1.2(3)\\
4 & 0.215(15)$\pm 0.805(15)i$ &1.00 &--- & --- & --- \\
1 & -0.22561(2)$\pm 0.68247(4)i$ &0.87 & -0.16(2) & 1.194(6) & 1.192(6) \\
2 & -0.394(5)$\pm 0.700(6)i$ &0.97 & --- & 1.8(6) & 1.6(4) \\
1 & -0.64890(4)$\pm 0.28696(4)$ &0.86 & -0.180(5) & 1.197(6) & 1.194(6) \\
3 & -0.685(15)$\pm 0.485(15)i$ &1.01 & --- & 2.3(5) & 1.4(3) \\
\mr \multicolumn{6}{c}{Spin-$\frac52$} \\ \mr
1 & 0.5501(3)$\pm 0.5842(2)i$ &0.91 & -0.4(1) & 1.19(2) & 1.19(4)  \\
3 & 0.522(5)$\pm 0.645(10)i$ &0.94 & -1.2(4) & & \\
1 & 0.0612(2)$\pm 0.7759(2)i$ &0.88 & -0.2(1) & 1.20(3) & 1.19(2) \\
3 & -0.03(1)$\pm 0.83(1)i$ &0.94 & --- & --- & --- \\
1 & -0.4227(1)$\pm 0.6400(1)i$ &0.87 & -0.20(5) & 1.185(15) & 1.21(3) \\
3 & -0.575(5)$\pm 0.61(2)i$ &0.95 & --- & --- & --- \\
3 & -0.665(15)$\pm 0.53(1)i$ &0.97 & --- & --- & --- \\
1 & -0.7213(2)$\pm 0.24595(15)i$ &0.87 & -0.175(25) & 1.20(2) & 1.20(2) \\
4 & -0.745(15)$\pm 0.39(2)i$ &0.96 & --- & --- & --- \\
\mr \multicolumn{6}{c}{Spin-3} \\ \mr
  & -0.92(1) &1.01 & --- & --- & --- \\
1 & 0.6608(4)$\pm 0.5232(5)i$ &0.93 & --- & 1.20(3) & 1.20(3) \\
3 & 0.645(15)$\pm 0.595(15)i$ &0.96 & -1.4(5) & 2.0(5) & 2.0(5) \\
1 & 0.2729(3)$\pm 0.7730(4)i$ &0.90 & --- & 1.20(4) & 1.19(4) \\
4 & 0.220(15)$\pm 0.840(15)i$ &0.95 & --- & 1.6(4) & 1.6(4) \\
1 & -0.1686(1)$\pm 0.7902(1)i$ &0.89 & -0.19(3) & 1.20(2) & 1.20(2) \\
2 & -0.275(5)$\pm 0.825(5)i$ &0.95 & --- & 1.2(3) & 1.2(3) \\
1 & -0.54955(5)$\pm 0.58351(3)i$ &0.88 & -0.20(4) & 1.196(6) & 1.197(5) \\
2 & -0.68(1)$\pm 0.54(1)i$ &0.95 & --- & 1.1(4) & 1.0(4) \\
1 & -0.76925(10)$\pm 0.21430(5)i$ &0.88 & -0.185(25) & 1.205(15) & 1.205(15) \\
\br
\endTable

\begin{figure}
\caption{\label{fig:sing} The distribution of singularities in 
the complex $u$-plane. In all cases the circle has radius $u_c$.}
\unitlength=.15mm
\begin{picture}(900,700)
\put(200,150){
\multiput(0,0)(350,0){3}{\multiput(0,0)(0,350){2}{
\put(-125,0){\vector(1,0){265}}
\put(0,-125){\vector(0,1){265}}
\multiput(0,-100)(0,100){3}{\line(1,0){10}}
\multiput(-100,0)(100,0){3}{\line(0,1){10}}
\multiput(0,-120)(0,20){13}{\line(1,0){5}}
\multiput(-120,0)(20,0){13}{\line(0,1){5}}
\put(-20,95){\scriptsize 1}
\put(94,-23){\scriptsize 1}
}}

\put(0,350){\circle{34.3}
\put(-105,105){\scriptsize $S=\frac12$}
\put(-100,0){\circle*{5}}}

\put(350,350){\circle{110.8}
\put(-105,105){\scriptsize $S=1$}
\put(-59.85,0){\circle*{5}}
\put(-30.19,37.88){\circle*{5}}
\put(-30.19,-37.88){\circle*{5}}}

\put(700,350){\circle{147.6}
\put(-105,105){\scriptsize $S=\frac32$}
\put(63,45){\circle*{5}}
\put(63,-45){\circle*{5}}
\put(9.48,64.12){\circle*{5}}
\put(9.48,-64.12){\circle*{5}}
\put(-6.55,71.1){\circle*{5}}
\put(-6.55,-71.1){\circle*{5}}
\put(-52.91,33.8){\circle*{5}}
\put(-52.91,-33.8){\circle*{5}}}

\put(0,0){\circle{165.9}
\put(-105,105){\scriptsize $S=2$}
\put(-85,0){\circle*{5}}
\put(37.7,64.0){\circle*{5}}
\put(37.7,-64.0){\circle*{5}}
\put(30.1,72.7){\circle*{5}}
\put(30.1,-72.7){\circle*{5}}
\put(21.5,80.5){\circle*{5}}
\put(21.5,-80.5){\circle*{5}}
\put(-22.6,68.2){\circle*{5}}
\put(-22.6,-68.2){\circle*{5}}
\put(-39.1,69.3){\circle*{5}}
\put(-39.1,-69.3){\circle*{5}}
\put(-64.9,28.7){\circle*{5}}
\put(-64.9,-28.7){\circle*{5}}
\put(-68.5,48.5){\circle*{5}}
\put(-68.5,-48.5){\circle*{5}}}

\put(350,0){\circle{175.9}
\put(-105,105){\scriptsize $S=\frac52$}
\put(54.9,58.5){\circle*{5}}
\put(54.9,-58.5){\circle*{5}}
\put(52.2,64.5){\circle*{5}}
\put(52.2,-64.5){\circle*{5}}
\put(6.1,77.6){\circle*{5}}
\put(6.1,-77.6){\circle*{5}}
\put(-3,83.0){\circle*{5}}
\put(-3,-83.0){\circle*{5}}
\put(-42.3,64.0){\circle*{5}}
\put(-42.3,-64.0){\circle*{5}}
\put(-57.5,61.0){\circle*{5}}
\put(-57.5,-61.0){\circle*{5}}
\put(-66.5,53.0){\circle*{5}}
\put(-66.5,-53.0){\circle*{5}}
\put(-72.1,24.6){\circle*{5}}
\put(-72.1,-24.6){\circle*{5}}
\put(-74.5,39){\circle*{5}}
\put(-74.5,-39){\circle*{5}}}

\put(700,0){\circle{182.2}
\put(-105,105){\scriptsize $S=3$}
\put(-92.3,0){\circle*{5}}
\put(66.1,52.3){\circle*{5}}
\put(66.1,-52.3){\circle*{5}}
\put(64.5,59.5){\circle*{5}}
\put(64.5,-59.5){\circle*{5}}
\put(27.3,77.3){\circle*{5}}
\put(27.3,-77.3){\circle*{5}}
\put(22.0,84.0){\circle*{5}}
\put(22.0,-84.0){\circle*{5}}
\put(-16.8,79.0){\circle*{5}}
\put(-16.8,-79.0){\circle*{5}}
\put(-27.5,82.5){\circle*{5}}
\put(-27.5,-82.5){\circle*{5}}
\put(-55.0,58.4){\circle*{5}}
\put(-55.0,-58.4){\circle*{5}}
\put(-68.0,54){\circle*{5}}
\put(-68.0,-54){\circle*{5}}
\put(-76.9,21.4){\circle*{5}}
\put(-76.9,-21.4){\circle*{5}}}
}
\end{picture}
\end{figure}
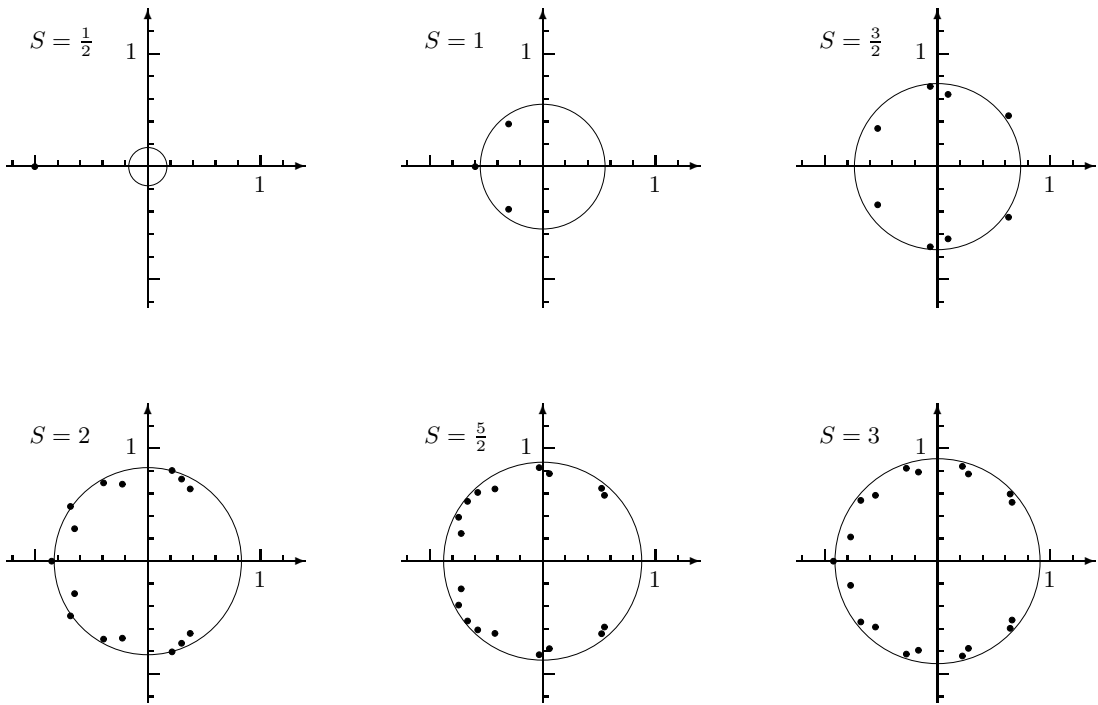

The distribution of singularities is shown in \fref{fig:sing}. A
remarkable feature of the singularity distribution is its 
regularity. As $S$ increase the complex singularities move closer
to the perimeter of the physical disc and the distance between
the various singularities become more uniform. In the limit
$S \rightarrow \infty$ it thus seems likely that the singularities
will converge onto the unit circle.

We find the very old conjecture by Fox and Guttmann \cite{fox}
that the number of singularities inside the physical disc
equals $qS-2$, where $q$ is the coordination number of the lattice
($q=4$ for the square lattice), to be
invalid for $S>1$. Recently, Matveev and Shrock \cite{MSISzeros}
studied the distribution of zeros of the partition function of 
the square lattice Ising model for $S= 1,\; \frac32$, and 2.  They
conjectured that all divergences of the magnetisation occur at 
endpoints of arcs of zeros protruding into the ferromagnetic phase 
and that there are $4[S^2]-2$ such arcs for $S \geq 1$, where $[x]$ 
denotes the integer part of $x$. Our analysis seems to confirm these 
conjectures for the magnetisation series up to $S=2$. In particular
we find evidence of singularities close to the endpoints located
by Matveev and Shrock \cite{MSISzeros} for these spin values.

The estimate for $\gamma '$ at the singularity $u_- = -1$ of the 
spin-$\frac12$ susceptibility and the estimates for the spin-1
series are based on the low-temperature series we published elsewhere 
\cite{enting94,JGS1}. The estimate for $\gamma '$ of the spin-$\frac12$
case is consistent with the exact value $\gamma ' = \frac32$ also 
reported by Matveev and Shrock \cite{MSISmodel}.

From \tref{tab:nps} we observe that the exponents at the singularities 
in the complex plane which are well converged (those marked 1)
appear to be independent of $S$. In 
the case of integer spin it appears that the exponents associated 
with the singularity on the negative $u$-axis equal those at $u_c$. 
While the exponents are independent of $S$, note that they do depend
on the lattice structure \cite{MSISmodel}, so a much weaker version
of universality holds at the non-physical singularities. In 
all these cases we observe that the Rushbrooke inequality 
\cite{rushbrooke},

\begin{equation}
  \alpha ' + 2\beta + \gamma ' \geq 2, \label{eq:rushbrooke}
\end{equation}
is satisfied, and it does indeed seem quite possible that the exponents
satisfy the equality in Eq.~(\ref{eq:rushbrooke}). At the remaining
singularities the errors on the exponent estimates are too 
large to make any such assertion.

\section*{E-mail or www retrieval of series}

The low-temperature series for the spin-$S$ Ising model can be obtained 
via e-mail
by sending a request to iwan@maths.mu.oz.au or via the world
wide web on http://www.maths.mu.oz.au/\~{ }iwan/ by following the
instructions.

\section*{Acknowledgements}

We would like to
thank Robert Shrock for his careful reading and
helpful comments on early versions of this paper.
Financial support from the Australian Research Council is gratefully 
acknowledged by IJ and AJG.


\section*{References}

\begin{thebibliography}{99}

\bibitem{enting94} Enting I G, Guttmann A J and Jensen I 1994
\JPA {\bf 27} 6987

\bibitem{fox} Fox P F and Guttmann A J 1973 \JPC {\bf 6} 913 

\bibitem{saul} Saul D M,  Wortis M and Jasnow D 1975 \PR B {\bf 11} 2571 

\bibitem{camp} Camp W J and Van Dyke J P 1975 \PR B {\bf 11} 2579

\bibitem{nickel} Nickel B G 1981 {\em Physica} {\bf 106A} 48 

\bibitem{MSISzeros} Matveev V and Shrock R 1995 \JPA {\bf 28} L533
  
\bibitem{sykes73} Sykes M F and Gaunt D S 1973 \JPA {\bf 6} 643

\bibitem{enting78} Enting I G 1978 \JPA {\bf 11} 563 

\bibitem{JGS1} Jensen I and Guttmann A J {\em Extrapolation procedure
for low-temperature series for the square lattice spin-1 Ising model}
submitted to J. Phys. A.

\bibitem{serana} Guttmann A J 1989 {\em Phase Transitions and Critical
Phenomena} vol 13, ed C Domb and J Lebowitz (New York:Academic) pp 1-234

\bibitem{LF} Liu A J and Fisher M E 1989 {\em Physica} {\bf 156A} 35

\bibitem{onsager} Onsager L 1944 \PR {\bf 65} 117 

\bibitem{yang} Yang C N 1952 \PR {\bf 85} 808 

\bibitem{IsSus} Barouch E, McCoy B M, and Wu T T 1973 \PRL {\bf 31} 1409 
  
\bibitem{MSISmodel} Matveev V and Shrock R 1995 \JPA {\bf 28} 1557

\bibitem{rushbrooke} Rushbrooke G S 1963 \JCP {\bf 39} 842 

\end{thebibliography}
\end{document}